\documentclass[prB, twocolumn, preprintnumbers ,amsmath, amssymb, superscriptaddress, aps, floatfix, longbibliography]{revtex4-2}

\usepackage{color}
\usepackage[T1]{fontenc}
\usepackage{amsmath,amssymb}
\usepackage{pifont}
\usepackage{amssymb}  
\usepackage{bbold}
\usepackage{float}
\usepackage{tikz}
\usepackage{caption}
\usepackage{subcaption}
\usepackage{makecell}
\usepackage{pifont}   
\usepackage{graphicx} 
\usepackage{dcolumn}  
\usepackage{bm}       
\usepackage{multirow} 
\usepackage[colorlinks=true, linkcolor=blue, citecolor=blue]{hyperref}
\usepackage{placeins}

\usepackage{graphicx}
\graphicspath{{figs/}}

\begin{document}
\title{Impact of clustering of substitutional impurities on quasiparticle lifetimes and localization}
\author{Jack G. Nedell}
\affiliation{Department of Physics, Northeastern University, Boston, MA 02115, USA}
\affiliation{Department of Physics, Cornell University, Ithaca, NY 14853, USA}
\author{Michael Vogl}
\affiliation{Department of Physics, King Fahd University of Petroleum and Minerals, 31261 Dhahran, Saudi Arabia}
\author{Gregory A. Fiete}
\affiliation{Department of Physics, Northeastern University, Boston, MA 02115, USA}
\affiliation{Department of Physics, Massachusetts Institute of Technology, Cambridge, MA 02139, USA}
\date{\today}

\begin{abstract}
Motivated by the observation and prediction of clustering behavior for impurities substituted into the host lattice of a real material, and the dramatic impact this can have on electronic properties, we develop a simple approach to describe such an effect via the electron self-energy. We employ a disorder averaged T-matrix expansion taken to second order, which we modify to include a clustering probability parameter. This approach circumvents the need for specific cluster probability distributions, simplifying greatly the analysis of clustered impurities. To gain analytical insights, we study a nearest-neighbor square lattice tight-binding Hamiltonian with clustered impurity substitutions to investigate clustering of off-diagonal hopping impurities.  We find that our T-matrix approach is in excellent agreement with exact numerical results from a tight-binding computation performed with the KWANT package.  We observe a variety of interesting impurity clustering-induced effects in the self-energy such as the suppression of quasi-particle lifetimes at certain momenta and an increase in localization, as indicated by the inverse participation ratio. The KWANT results are reproduced in our modified T-matrix approach. In addition, our method allows for a full analytical treatment of clustering effects which can aid in physical insight. 
\end{abstract}

\maketitle

\section{Introduction}

The interplay between disorder and the electronic properties of two-dimensional (2D) materials \cite{zou_open_2015} is of great interest. Of particular importance are the questions of how to control and minimize undesirable disorder effects, as well as how to harness disorder to achieve desirable physical properties (e.g., quantized responses such as occurs in the quantum Hall effect or quantum anomalous Hall effect).

An interesting yet underappreciated aspect of disorder in materials is the spatial configuration of impurities, which most often is not taken into account since a disorder average is typically employed. In a given material, do impurities cluster or are they well-separated? In general it is non-trivial to find the energetically most favorable configuration of impurities, whether they are adsorbed atoms on a material surface or chemically substituted within a host material lattice. Various density functional theory (DFT) studies have been performed to address this question for numerous host materials and impurity types--for instance, in studies that investigate impurities substituted \cite{kullgren_dft-based_2017} or adsorbed \cite{forster_phase_2012} onto the surface or the bulk \cite{furthmuller_clustering_2012} of materials. Often one finds that impurities show preference for clustering with clusters of certain preferred sizes. The details of this effect depend on the host material, the impurity type, temperature, and the concentration of impurities.

Likewise, similar clustering behavior has been observed or induced experimentally. For instance, adatoms on the surface of graphene, have been observed to emerge in clusters \cite{wang_electronic_2015,forster_phase_2012}, and a recent study demonstrated the formation of large nitrogen clusters substituted into the graphene lattice, with a reported increase in conductivity \cite{lin_nitrogen_2019}.

There have also been theoretical studies of substitutional impurities in graphene \cite{berdiyorov_effect_2016,peres_electronic_2006,peres_electron_2007,pereira_modeling_2008,javan_theoretical_2020} and studies of the effect of impurity clusters \cite{sule_clustered_2014,katsnelson_scattering_2009} on electronic properties. Substitutional impurities in materials are known to harbor localized states \cite{anderson_localized_1961}, which for graphene \cite{peres_electron_2007} has been suggested to play an important role in the transport properties of nanoribbons as transport varies dramatically with specific disorder configuration, relative to the edges of the system \cite{berdiyorov_effect_2016}. Another interesting effect is a recently predicted metal-to-insulator transition in disordered graphene \cite{kostadinova_spectral_2019}.

Finally, it is important to point to the theory of random binary alloys, which received heightened attention in the 1970s \cite{mills_analytic_1978,diehl_pseudofermion_1979}. Such studies were done in the framework of the coherent potential approximation (CPA). This previous work expanded upon details of scattering by large clusters of impurities, including impurities with off-diagonal disorder, which will be our focus our focus as well. These past and present research efforts set the stage for the investigation we present here. 

Using a T-matrix expansion, we compute the self energy of a nearest neighbors tight-binding model on a square lattice with off-diagonal (hopping) impurities. We compare the results for an ordinary T-matrix description of isolated impurities to our  T-matrix model, which allows us to take into account clustering of neighboring impurities. We pit these results against a disorder-averaged self-energy obtained from exactly solved finite systems realizing the same disorder model. We find that clustering contributes non-trivially to the self-energy, especially the imaginary part. We find our T-matrix approach is in excellent agreement with exact numerical results. Yet, it requires almost no additional technical know-how beyond the standard T-matrix approach and thus is of great practical utility.

We proceed to use this self-energy to understand how electronic properties of the system vary with the concentration of impurity clustering. In particular, we compute the density of states, quasi-particle lifetime, and inverse participation ratio, which gives us an indication of localization properties. This analysis complements the results of previous studies, demonstrating increased localization and a decrease in the zero-energy density of states for increased clustering.  These results are summarized in the figures of the text.

Our paper is organized as follows.  In Sec.~\ref{sec:model} we introduce the tight-binding model, including disorder, on the square lattice that we will use for our study of impurity clustering.  In Sec.~\ref{sec:self_energy} we introduce the T-matrix approach and present analytical results on isolated and clustered impurities. In Sec.~\ref{sec:clustering} we present numerical results on disorder effects related to the clustering of impurities. Finally, in Sec.~\ref{sec:conclusions} we present the main conclusions of our work.

\section{Model}
\label{sec:model}
As a platform for our investigation of impurity clustering effects we choose a tight-binding model on a square lattice. This model consists of only nearest-neighbor hoppings. This choice of model has the advantage that much analytic progress is possible and will therefore allow for deeper insights into our treatment of impurity clustering.  However, our main results are more general and do not depend on this specific Hamiltonian or lattice.  

The Hamiltonian of the clean lattice may be written as a sum over the lattice sites of the model $\bf r_i$, and the nearest-neighbor displacement vectors $\boldsymbol \delta$,
\begin{equation}
\hat{H}_0 = -t \sum_{\mathbf{r}_i,\boldsymbol\delta} \hat{c}^\dagger(\mathbf{r}_i)\hat{c}(\mathbf{r}_i+\boldsymbol\delta).
\end{equation}
Fourier transforming to a crystal momentum basis one obtains the dispersion relation $\xi_\mathbf{k} = -2 t (\cos(k_xa)+\cos(k_ya))$, where $a$ is the lattice spacing. We take $t=1$ and $a=1$. 

Next, we introduce to this model a set of lattice site ``substitutions". If the original lattice is made up of atoms of type A, and we replace select lattice sites with atoms of type B, hoppings between atoms are no longer only $t=t_{AA}$ (hopping among A atoms) but may also be $t_{AB}=t_{BA}$ (hopping between A and B atoms) or $t_{BB}$ (hopping among B atoms). The full Hamiltonian we consider is $\hat{H} = \hat{H}_0 + \hat{H}_{imp}$, where $\hat{H}_{imp}$ is the impurity Hamiltonian. We write the impurity Hamiltonian in terms of the set of impurity coordinates, $\{ \mathbf{R}_i\}$, and the hopping corrections $ t' = t - t_{AB}$ and $t'' = \frac{1}{2}(t+t_{BB}-2t_{AB})$,
\begin{equation}
\hat{H}_{imp} = \sum_{\mathbf{R}_i,\boldsymbol\delta} \Delta t (\mathbf{R}_i, \boldsymbol\delta)\left(\hat{c}^\dagger(\mathbf{R}_i)\hat{c}(\mathbf{R}_i+\boldsymbol\delta) + h.c. \right),
\label{eq:himp}
\end{equation}
with the hopping term $\Delta t(\mathbf{R}_i,\boldsymbol\delta) =  t' -  t'' \sum_{\mathbf{R}_j} \delta(\mathbf{R}_i+\boldsymbol\delta - \mathbf{R}_j)$ replacing hoppings appropriately. In the standard treatment of impurities which implicitly assumes they do not neighbor each other on the lattice, we have simply $\Delta t (\mathbf{R}_i,\boldsymbol\delta) = t'$, {\it i.e.} a dependence of the neighboring lattice positions does not enter the hopping correction and only one sum over the impurity coordinates is needed in the impurity Hamiltonian. Note that even in a ``simpler" case of $t_{AB}=t_{BB}$, this additional sum over lattice sites and the associated terms in the T-matrix expansion are still needed to correctly account for neighboring impurities.

\section{Self-Energy Computation}
\label{sec:self_energy}
We study the effect of the impurities on the self-energy appearing in the single-particle Green's function.  The imaginary part will reflect electron lifetime effects, and the real part will reflect energy renormalizations.

\subsection{T-matrix approach}
In this work we make use of the disorder-averaged T-matrix. It can be written in terms of the impurity Hamiltonian and the undressed square lattice Green's function $G_0(i\omega_n,\mathbf{k}) = \frac{1}{i\omega_n - \xi_\mathbf{k}}$ (where $\omega_n$ are the Matsubara frequencies), as an expansion in the hopping correction \cite{altland_condensed_2010}:
$$\hat{T} = \langle \hat{H}_{imp} + \hat{H}_{imp} \hat{G}_0 \hat{H}_{imp} + ... \rangle_{dis}.$$
Here, the disorder average $\langle ... \rangle_{dis}$ consists of an integration over the impurity coordinates, divided by the total area, {\it i.e.} an average over a uniform distribution. Averaging over a nonuniform probability distribution for the impurity locations would require more care, but we circumvent this in our discussion of impurity clusters.
In what follows we will consider first the case of isolated impurities, which follows a standard treatment. Afterwards we consider the case of clustered impurities, which we are able to treat semi-heuristically via the introduction of a clustering probability.

\subsubsection{Isolated Impurities}
Operating under the assumption that impurities are isolated, {\it i.e.} impurities do not occupy neighboring sites, the self-energy can be taken as the number of impurities times the disorder averaged T-matrix for a single impurity, $\Sigma_\mathbf{kk'} = N_{imp} T_\mathbf{kk'}$. The infinite T-matrix series can be summed analytically, which we discuss in Appendix \ref{sec:single}. From this series, the T-matrix self-energy for isolated hopping impurities on the square lattice, in terms of the dimensionless hopping correction $t'$ and the impurity concentration $\sigma = \frac{N_{imp}}{N}$ is given by
\begin{equation}
\Sigma_T(\omega,\mathbf{k}) =  \frac{-t'\sigma}{d(\omega)} \left(a(\omega) \xi^2_\mathbf{k}+ b(\omega) \xi_\mathbf{k}+c(\omega) \right),
\end{equation}
where we have defined the coefficients as 
\begin{eqnarray}
        a(\omega) &=& t' g(\omega), \\
        b(\omega) &=& -2(1+t'(\omega g(\omega)-1)),\\
        c(\omega) &=& \omega t'(\omega g(\omega)+1), \\ 
        d(\omega) &=& t'^2(\omega g(\omega)-1) - 2t'(\omega g(\omega) -1) -1.
\end{eqnarray}

\begin{figure}[H]
    \centering
    \includegraphics[width=8cm]{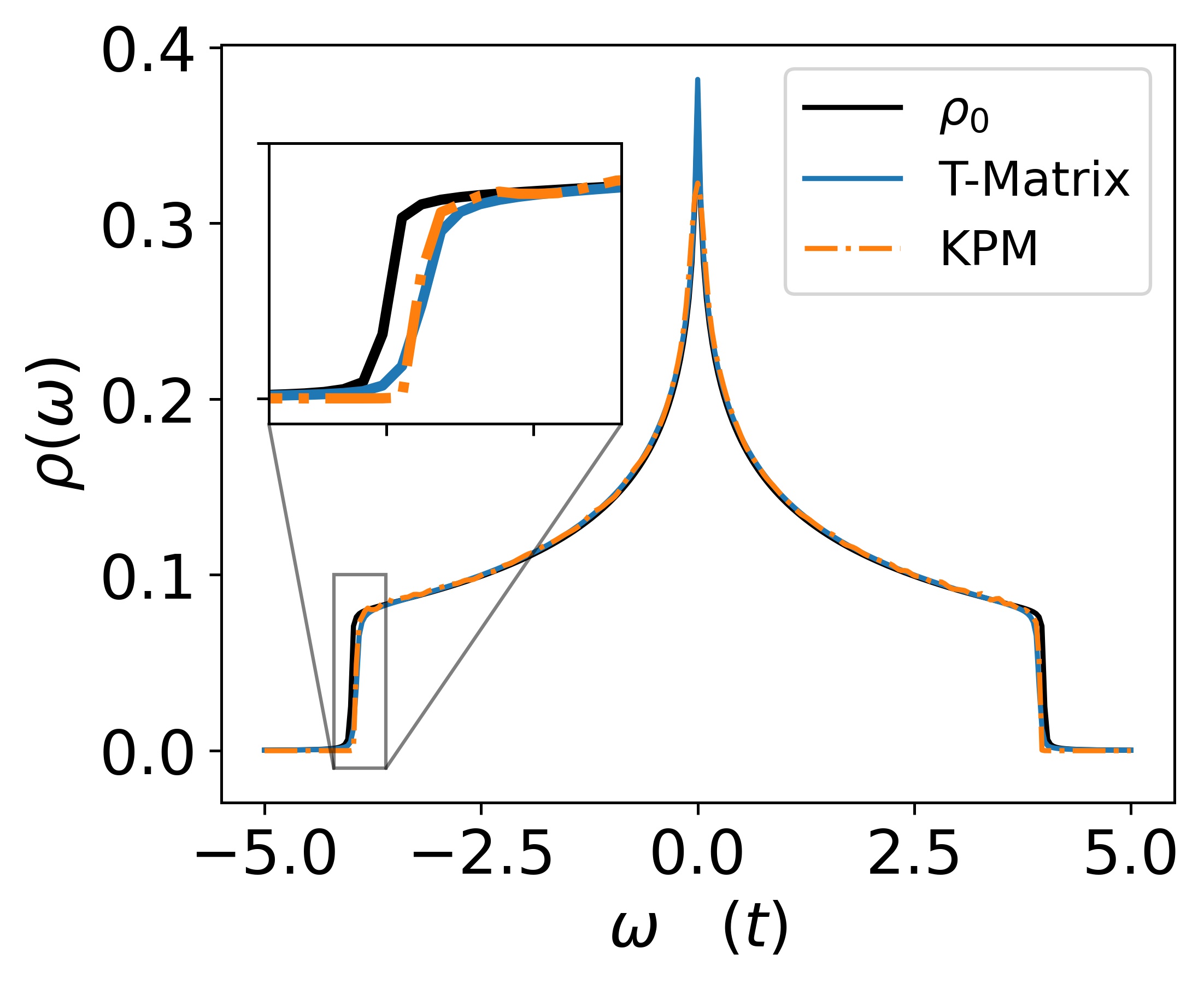}
    \caption{Density of states, for zero disorder ($\rho_0$) and for disorder at a concentration of $\sigma=0.1$. Hopping parameters used are $t=1$ for the hoppings in the clean system, $t_{AB}=0.9$ and $t_{BB}=1.1$ for hoppings to impurity substitutions and between impurities. In the plot, we compare the kernel polynomial method (KPM) expansion on a finite model system of 401x401 lattice sites, which was computed in KWANT \cite{groth_kwant_2014} with our T-matrix result (integrals have been obtained with the QUADPY package \cite{schlomer_nschloequadpy_2021}). Details of the model implementation for the finite system are discussed in Sec. \ref{numeric}.}
    \label{fig:dos}
\end{figure}

The form of this T-matrix based self-energy is similar to that obtained previously for an isolated impurity in graphene \cite{peres_electron_2007}.
The term $g(\omega)$ appearing in the expressions above is given by $g(\omega) = \int_{B.Z.} \frac{d^2 k}{(2 \pi)^2} G_0(\omega,\mathbf{k})$, the integral of the undressed square lattice Green's function over the first Brillouin zone. The integral may be evaluated as an elliptic integral of the first kind, written in modulus form \cite{kogan_greens_2021},
\begin{equation}
     g(\omega) = \int_{-\pi}^{\pi}\int_{-\pi}^{\pi} \frac{d\mathbf{k}}{(2\pi)^2} \frac{1}{\omega +i\eta - \xi_\mathbf{k}} = \frac{2}{\pi \omega} \mathbf{K}(\frac{4t}{\omega}),
     \label{eq:sq}
\end{equation}
where the Matsubara frequency has been analytically continued to real frequecies as $\omega_n\to \omega+i\eta$ with $\eta=0^+$.
An interesting and  measureable quantity that can be found from the T-matrix self-energy is the density of states (DOS), which we have plotted in Fig. \ref{fig:dos}.

From Fig.\ref{fig:dos} we find that there is good agreement between the DOS as computed with the isolated-impurity T-matrix and the random-impurity numerical result obtained in KWANT \cite{groth_kwant_2014} with the Kernel Polynomial Method (KPM). It should be stressed that we have this good agreement despite the fact that our numeric computation allows for neighboring impurities with the different hopping parameter $t_{BB}$. This is in contrast to the T-matrix result, which assumed isolated impurities. Our result demonstrates the relatively wide range of applicability of the isolated impurity assumption for random uniform distributions of impurities. This agreement is intuitive, as at low concentration in random uniform impurity distributions it is unlikely for impurities to neighbor one another. 

One main difference between the computed DOS can be observed at $\omega=0$, where the KPM result is suppressed compared with the isolated-impurity T-matrix result. Later analysis with our clustered impurity method will show that while this type of suppression does occur due to clustering of impurities, the magnitude of the suppression is larger than would be expected from impurity clustering alone, such that this may be attributed to nature of the numerical calculation, likely a finite size effect or an effect of the KPM.

\subsubsection{Clustered Impurities}
To allow impurities to neighbor one another in our model, we must work with the full impurity Hamiltonian in Eq.(\ref{eq:himp}). We take the first two terms in the T-matrix series expansion, which correspond to one and two-fold scattering from a single impurity. Terms corresponding to multiple-fold scattering from multiple distinct impurities or distinct clusters of neighboring impurities are discarded in accordance with the T-matrix approximation. Full summation of this T-matrix series, which at order $n$ describes $n$ fold scattering by a single impurity or by up to $n$ neighboring impurities in a cluster, is a much greater task in our case than in the case for isolated impuritites. Thus, we restrict ourselves to the first two terms. We note that the reason for this is that the first nonvanishing imaginary contribution to the self-energy appears at quadratic order in the hopping corrections and we therefore want to capture the impact of this contribution. For an analytic approach to systems where large impurity clusters and many-fold scattering are important, CPA methods \cite{mills_analytic_1978,diehl_pseudofermion_1979} would be preferred.

To model clustering of impurities, we employ a semi-heuristic approach. The T-matrix includes terms $\propto\sigma^2$, where $\sigma$ is the probability of placing a impurity at a given lattice site. We may then interpret $\sigma^2$ as the probability that two impurities sit next to each other. If we now replace these $\sigma^2$ appropriately to involve a new probability $p$, which represents a nonspecific clustering of impurities, we find that the effect due to neighboring impurities can be captured semi-heuristically this way. More details will be provided for our example below. This is an intuitive albeit heuristically implemented approach, which enables us to very simply generalize the T-matrix to a nonuniform impurity distribution.  It also agrees rather well with exact numerics as we show below.

Let us now illustrate for our example how such a substitution is made by explicitly evaluating  the first order term in the T-matrix expansion, $\langle H_{imp} \rangle_{dis}$. This begins with evaluating the term in the random-uniform distribution of impurities. Expressing this in terms of $\langle \mathbf{k} | H_{imp} |\mathbf{k}'\rangle $,
\begin{multline*}
    T^{(1)} = \frac{1}{N^2} \sum_{\mathbf{R}_i,\mathbf{R}_j,\boldsymbol\delta}(e^{i \mathbf{k'}\cdot \boldsymbol\delta} + e^{-i \mathbf{k}\cdot \boldsymbol\delta}) \times \\ \int d^2R_i d^2 R_j (t' - t''\delta(\mathbf{R}_i+\boldsymbol\delta - \mathbf{R}_j))e^{i\mathbf{R}_i\cdot(\mathbf{k}'-\mathbf{k})} \\
    = \delta_{\mathbf{k},\mathbf{k}'}\frac{-2\xi_\mathbf{k}}{N^2}(NN_{imp}t' - N_{imp}^2t'') = -2\xi_\mathbf{k}\delta_{\mathbf{k},\mathbf{k}'}(\sigma t' - \sigma^2t''),
\end{multline*}
where the dependence on the number of impurities $N_{imp}$ and the number of total sites in the system $N$ may be reduced to their ratio, the impurity concentration $\sigma$. 

The generalization to impurity clustering occurs with terms of order $\sigma^2$ and greater. In the term above, for example, $\sigma^2$ represents the probability that any one site and it's neighbor in any one direction both host an impurity. In the random-uniform distribution of impurities, these impurity occupancy probabilities are independently $\sigma$, leading to a proportionality with $\sigma^2$. We are interested, however, in clustered distributions, where impurity occupancy on a site and impurity occupancy of neighboring sites are not, a priori, independent. To model this, we keep the probability for impurities to occupy isolated sites as $\sigma$, but in the correction terms which correspond to the case of neighboring impurities, we assume that occupancies of these sites neighboring an existing impurity all occur independently with variable probability $\frac{p}{4}$. Using this simple scheme, the above term would be modified as $-2\xi_\mathbf{k}(\sigma t' - \sigma^2t'') \rightarrow -2\xi_\mathbf{k}(\sigma t' - \frac{\sigma p}{4}t'')$.

Applying this modification to the impurity averaged T-matrix at second order, $T = \langle H_{imp} +  H_{imp}G_0H_{imp}\rangle_{dis}$, the self energy may be written as follows,
\begin{multline}
    \Sigma(\omega,\mathbf{k}) =  -\alpha \xi_\mathbf{k} + \\ \int_{B.Z.}\frac{d^2 q}{(2\pi)^2}  \left[  \beta(\xi_\mathbf{k}+\xi_\mathbf{q})^2 - \gamma (\xi_\mathbf{0}+\xi_\mathbf{k+q})\right]G^0_\mathbf{q},
    \label{eq:Sneighb}
\end{multline}
with
\begin{eqnarray}
\label{eq:Tcoeff}
    \alpha &=& 2 \sigma t' - \frac{\sigma p}{2} t'',  \\
    \beta  &=&\sigma (t')^2 - \sigma p t't'' + \frac{\sigma p^2}{4}(t'')^2, \\
    \gamma &=& \frac{1}{2}\sigma p t (t'')^2\;.
    \label{eq:Tcoeff_final}
\end{eqnarray}
As it turns out, the integral involved in Eq.\eqref{eq:Sneighb} is again tractable in terms of elliptic integrals. We introduce the function $h(\omega)$, which is Eq.(3.148) in Ref. \cite{gradshtein_table_2007},
\begin{multline}
h(\omega) = \int_{B.Z.} \frac{d^2q}{(2\pi)^2} \frac{\operatorname{cos}(q_x)}{\omega+i\eta-\xi_\mathbf{q}}\\ 
= \frac{2}{\pi \omega}\left(\left(\frac{\omega}{t}+2\right)\boldsymbol\Pi\left(-\frac{4t}{\omega},\frac{4t}{\omega}\right) - \left(\frac{\omega}{t}+1 \right)\mathbf{K}\left(\frac{4t}{\omega}\right) \right).
\end{multline}
where $\mathbf{K}(k)$ and $
\boldsymbol\Pi(n,k)$ are the complete elliptic integrals of the first and third kind, respectively, where an analytic continuation $\omega = \omega + i \eta$ is implicit. We also have kept the hopping parameter $t$ symbolic in this expression to make the units clear, although as mentioned we set $t=1$ for our numerics. In terms of $g(\omega)$ and the function $h(\omega)$, we can express the self energy as
\begin{multline}
\label{eq:analytic}
\Sigma(\omega,\mathbf{k}) = -\alpha\xi_\mathbf{k} + \beta \left((\omega + \xi_\mathbf{k})^2 g(\omega)-2\xi_\mathbf{k}-\omega \right)\\ - \gamma\left( \xi_\mathbf{0} g(\omega) + \xi_\mathbf{k} h(\omega)\right).
\end{multline}
The analytical result, Eq.\eqref{eq:analytic}, is the main result of this section.  It is exact at the level of approximation at which we are working and contains the information about the clusters of the impurities through Eq.\eqref{eq:Tcoeff}-Eq.\eqref{eq:Tcoeff_final}.

\subsection{Exact numerical treatment of clustered impurities} \label{numeric}

For a more complete investigation of the physics at hand, and to corroborate the correctness of our treatment of impurity clustering, we also compute the self-energy of exact numerical realizations $H$ of the Hamiltonian in Eq.(\ref{eq:himp}). The Green's functions are computed as $G = \left((\omega+i\eta)\mathbb{1} - H \right)^{-1}$, where $\omega$ is the energy and $\eta$ a small numerical factor that was inserted for convergence, chosen as $\eta=0.01$.

In our computations we considered a lattice with $N=61\times 61$ sites. Inside the lattice we placed $N_{imp}$ impurities
, which corresponds to an impurity concentration $\sigma=N_{imp}/N$. Of the $N_{imp}$ impurities a percentage $p$ (called clustering probability) was generated such that they are part of a cluster of impurities with random size. The remaining $(1-p)N_{imp}$ impurities were placed randomly but isolated from other impurities or clusters. This algorithm for generating impurity distributions is non-specific in the choice of cluster sizes in the same way that our analytic T-matrix approach is, {\it i.e.} there is no preference for smaller or larger clusters. There is merely some fixed probability that each impurity neighbors another. This can also be expected to be true in various experimental scenarios, which is part of the reason that we consider this case.
\begin{figure*}[htbp!]
    \centering    \includegraphics[width=0.7\textwidth]{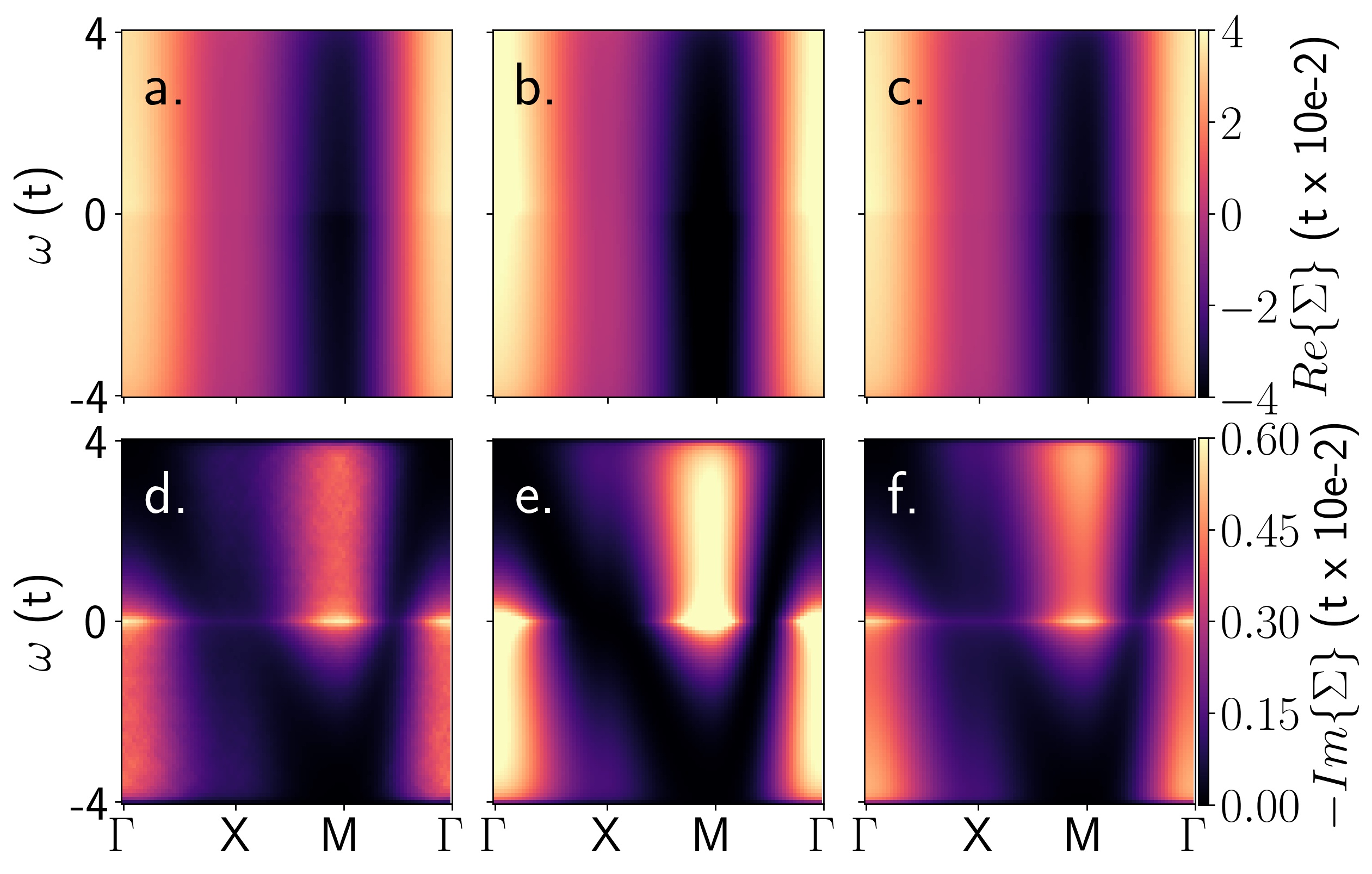}
    \caption{{\bf The self-energy.} Real part (top row) and imaginary part (bottom row). Column 1 (a,d) are the result of the numerical averaging of finite model Green's functions, while column 2 (b,e) and column 3 (c,f) are the T-matrix results neglecting and including neighboring impurities, respectively. These plots are computed for parameters $(\sigma, t_{AB},t_{BB}) = (0.05, 0.9, 1.1)$ at an impurity clustering of $p=0.3$}
    \label{fig:Seng}
\end{figure*}

To find the disorder averaged self-energy we applied the following numerical procedure. For a desired concentration of $\sigma$ and clustering probability $p$, a disorder-averaged Green's function was obtained by averaging $n_{ave}=20$ disorder realizations. The averaged self-energy in position space was then obtained as $\Sigma = (G_0)^{-1}-\langle G \rangle_{avg}^{-1}$. We then transformed to an approximate crystal momentum basis with the discrete Fourier transform matrix,
\begin{equation}
U_{ij} = \frac{1}{\sqrt{N}}e^{i\mathbf{k}_i\cdot \mathbf{r}_j},
\end{equation}
where the self-energy was obtained as $\Sigma_{\mathbf{kk'}}=U^\dagger \Sigma_{\mathbf{rr'}}U$. The diagonal components, which dominate the numerically averaged self-energy, were then extracted for comparison with the T-matrix method. Off diagonal elements correspond to scattering between crystal momentum eigenstates. While these elements may be relevant in a single disorder realization, they vanish analytically in the disorder average framework, and likewise are observed to vanish in our numerical disorder average. The motivation for this disorder average is self-averaging systems, as off-diagonal scattering is likewise negligible in these very large systems with homogeneous disorder.

\section{Effects of Impurity Clustering}
\label{sec:clustering}

The exact numerical computation mentioned above provides us with an approximate self-energy for self-averaging systems hosting hopping impurities, characterized by impurity density $\sigma$ and clustering probability $p$. We now may investigate how clustering impacts the physics of the model. We perform this analysis with the same parameters, $t_{AB}=0.9$ and $t_{BB}=1.1$ as before. We do not discuss different values of $(t_{AB},t_{BB})$ because for the regime where our approach is applicable the self-energy $\Sigma\propto (t'')^2$, which tells us that a change in parameters will mostly just increase the strength of an observed effect but not its nature.

First, we compare the self energy for various approaches in Fig.~\ref{fig:Seng}. Note that for the purpose of visual clarity we have used $\eta=0.05$ rather than $\eta=0.01$ in the plot for the self energy.  In the perturbative regime, when $t_{AB}$ and $t_{BB}$ are both reasonably close to the original hopping $t$, we observe that the main effect of impurity clustering is imprinted on the imaginary part of the self-energy. In the real part it primarily leads to a renormalization of the bandwidth. It is dominated by the term at first order in the hopping corrections, which in the perturbative regime has a relatively small effect due to impurity clustering.

\begin{figure}[htbp!]
    \centering
    \includegraphics[width=8cm]{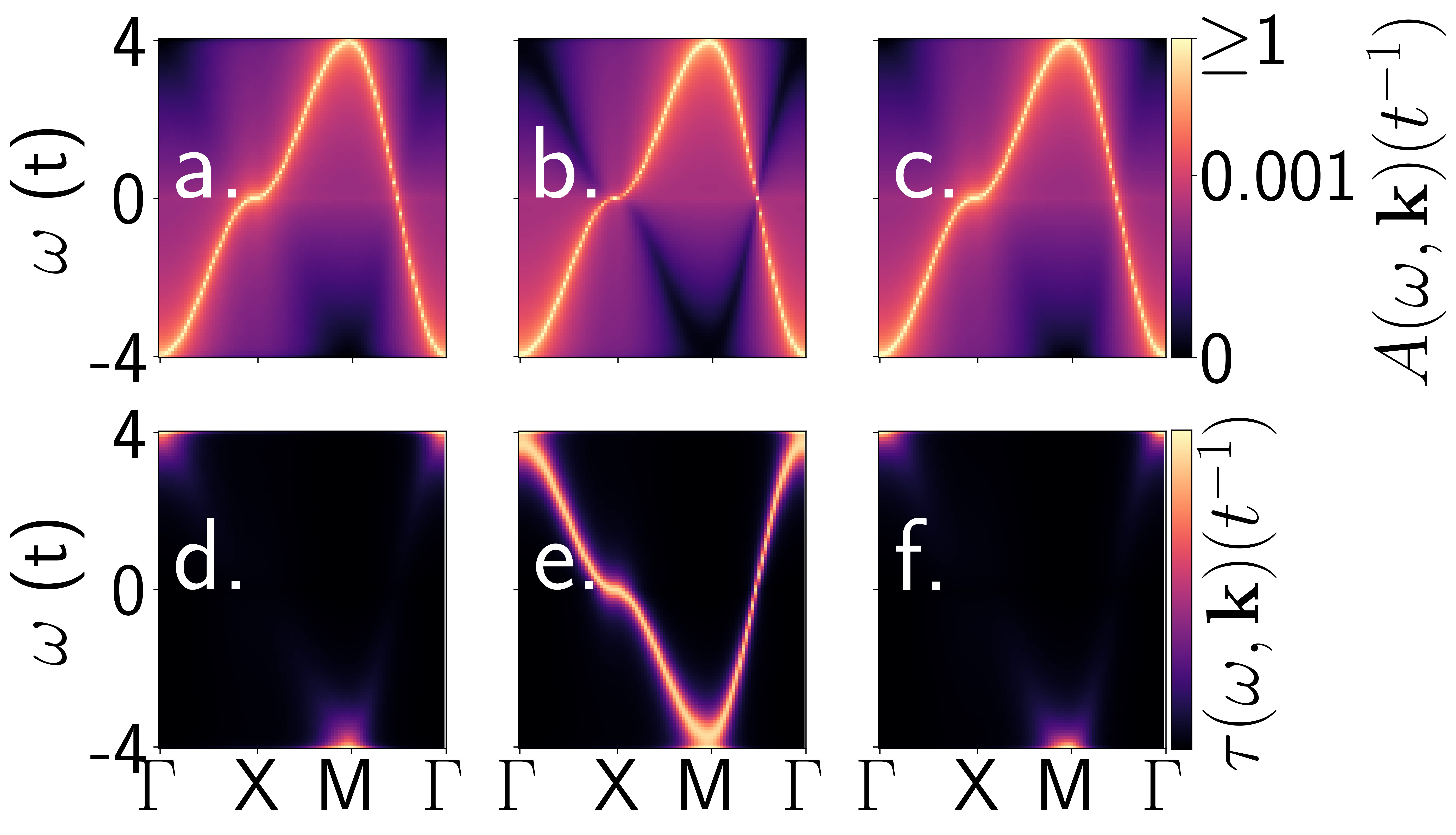}
    \caption{Spectral function (top) and quasiparticle lifetime (bottom) for the three computation methods. Column 1 (a,d) are the result of the numerical averaging of finite model Green's functions, while column 2 (b,e) and column 3 (c,f) are the T-matrix results neglecting and including neighboring impurities, respectively. These plots are computed for parameters $(\sigma, t_{AB},t_{BB}) = (0.05, 0.9, 1.1)$ at an impurity clustering of $p=0.3$}
    \label{fig:Atau}
\end{figure}

Next, in Fig.~\ref{fig:Atau} we plot the spectral function and quasiparticle lifetime, computed from the self energy as $A(\omega,\mathbf{k}) = -\frac{1}{\pi} \operatorname{Im} \{G(\omega,\mathbf{k})\}$ and $\tau(\omega,\mathbf{k}) = \frac{-1}{2 \operatorname{Im}\{\Sigma\}}$.  We find that the spectral function $A(\omega,\mathbf{k})$, much like the real part of the self-energy, changes very little due to impurity clustering. However, the quasi-particle lifetime is significantly different. Clustering of impurities here leads to significant suppression of quasi-particle lifetimes for most of the energy range.

\begin{figure}[htbp!]
    \centering
    \includegraphics[width=8cm]{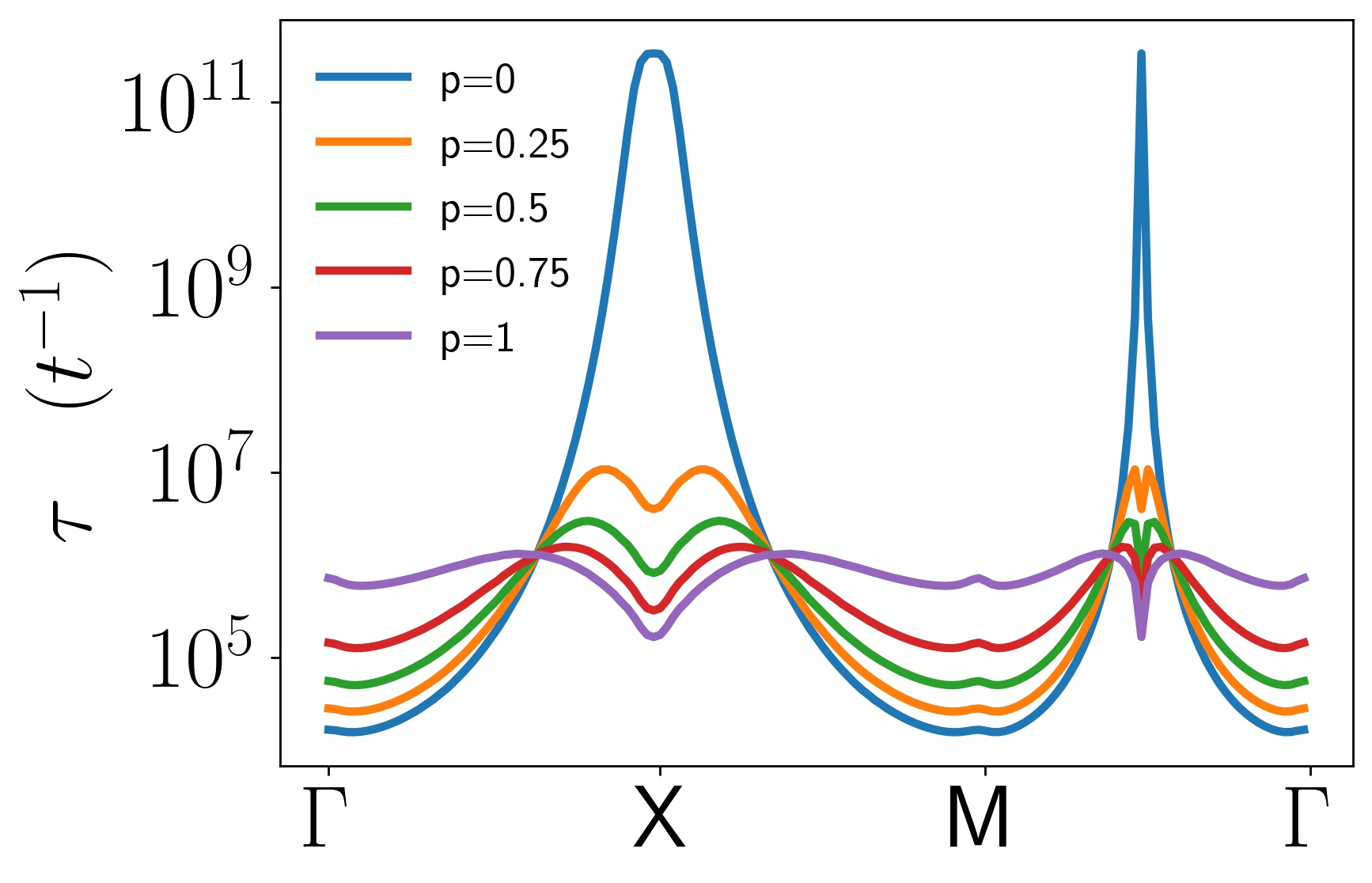}
    \caption{Quasiparticle lifetime along renormalized band $\xi'_\mathbf{k}$, with varying cluster parameter $p$.}
    \label{fig:tau}
\end{figure}

We can visualize this effect even better, if we examine the quasiparticle lifetime along the renormalized energy band of the model, $\tau(\xi'_\mathbf{k},\mathbf{k})$. In Fig.~\ref{fig:tau} we have plotted the quasi-particle lifetime for different values of the clustering probability $p$.  Because the energy band of the model along which we want to compute the lifetime is renormalized by the disorder, we take this renormalization from the first order of the T-matrix, $\xi'_\mathbf{k} = (1+2\sigma t'-\frac{\sigma p}{2}t'')\xi_\mathbf{k}$, which is the dominant real part of the self-energy and thus the dominant contribution to the rescaling of the energy band. We observe that with increasing clustering probability $p$, there is a general flattening of the lifetime. Maxima, previously located between symmetry points $M$ and $\Gamma$ and at $X$- that is specifically at the points where $\xi_\mathbf{k}=0$, are significantly flattened and we observe them splitting into two peaks. This effect does not depend on whether $t_{BB}$ is larger or smaller than $t$, as the term responsible for these effects is $\propto (t'')^2$ and therefore only depends on the difference.

\begin{figure}[t]
    \centering
\includegraphics[width=8cm]{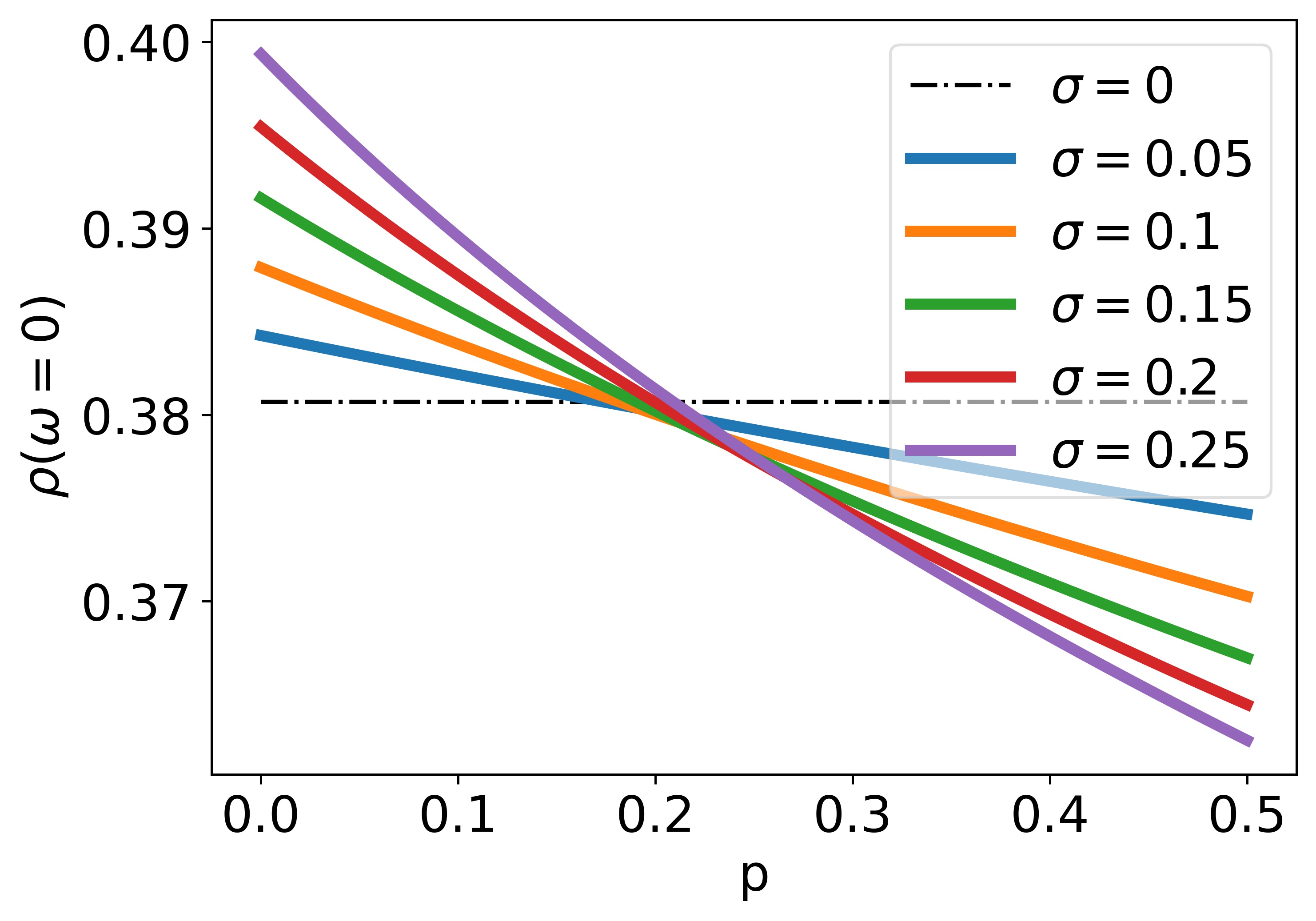}
    \caption{Zero-energy density of states for different impurity concentrations as they vary with the clustering parameter.}
    \label{fig:dos0}
\end{figure}

We now return to an observation made previously during our discussion of the isolated impurity approximation: at $\omega=0$ the numerically computed DOS with random neighboring impurities was smaller than the value expected from the analytical result within the isolated impurity approximation. To investigate this further, and determine whether this may be a clustering-related effect, we numerically integrate the Green's function as corrected with the impurity cluster self energy at $\omega=0$. The results are shown in Fig.~\ref{fig:dos0}.
We  observe that increasing clustering (larger clustering probability $p$) leads to a decrease in the density of states at $\omega=0$. This decrease is again attributable to the term in the self-energy $\propto (t'')^2$, and will be observed with any values $t_{AB}$ and $t_{BB}$. However, the magnitude of the decrease predicted to occur for larger clustering of impurities is smaller than the difference observed between the DOS curves in Fig. \ref{fig:dos}, suggesting that this particular discrepancy is related to the difference in the computation method rather than representative of interesting physics.

Finally, we consider a long-standing feature of disorder models \cite{anderson_localized_1961}, localization. One may capture numerically how localized the states in a given physical system are with the so-called inverse participation ratio (IPR).  To get an overall measure of the level of localization present in our system we sum the IPR for all states as below,
\begin{equation}
    \operatorname{IPR} = \sum_{i,j} |\psi_j(\mathbf{x}_i)|^4.
\end{equation}

\begin{figure}[h]
    \centering
    \includegraphics[width=8cm]{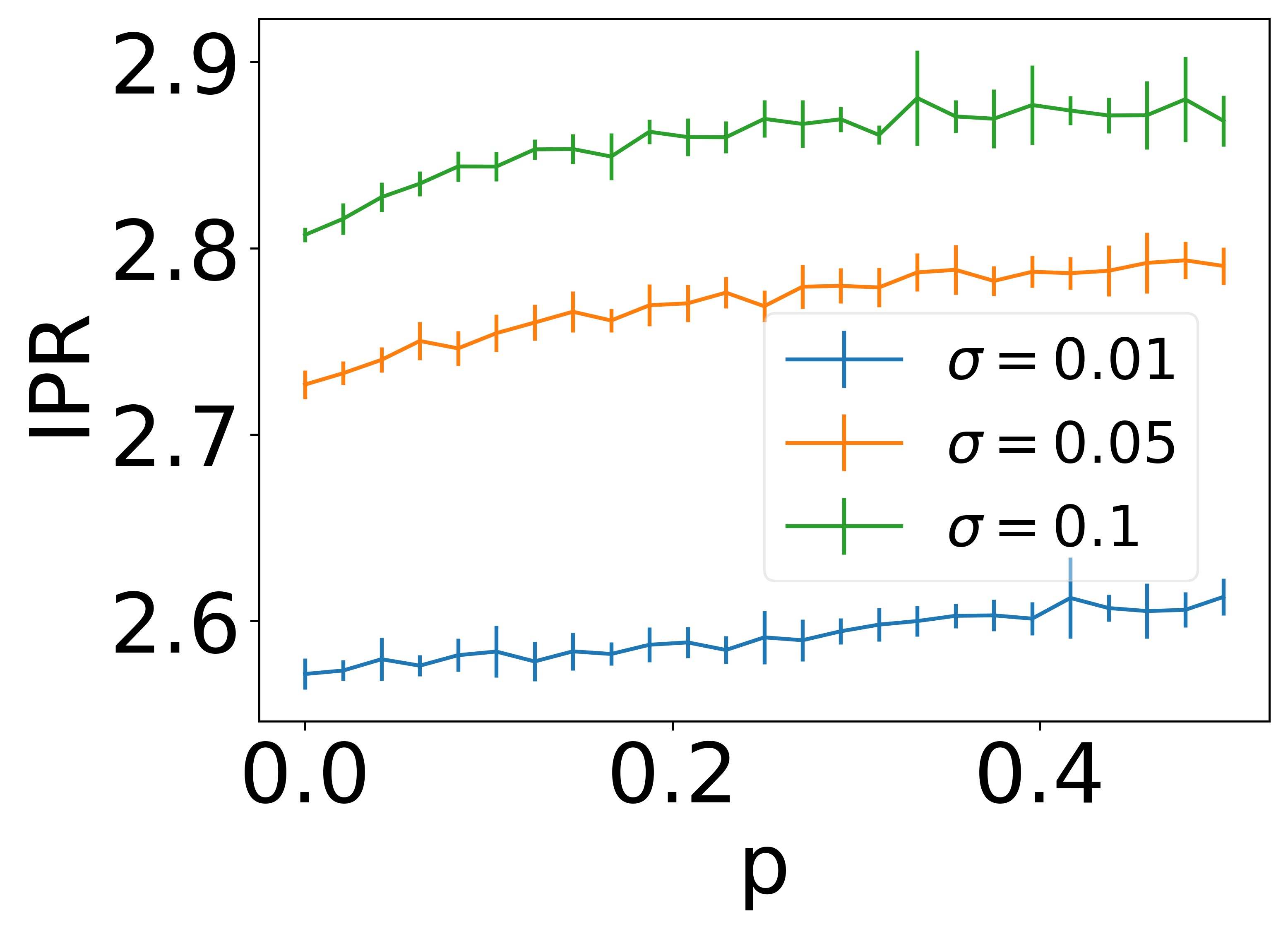}
    \caption{Inverse participation ratio (IPR) varying with disorder concentration and clustering, computed numerically with an average of 10 systems. Larger IPR indicates more localized states. Error bars show standard deviation of numerical average.}
    \label{fig:ipr}
\end{figure}

To get a result valid for various impurity configurations we average the inverse participation ratio (IPR) over 10 realizations of the system. In Fig.~\ref{fig:ipr} we have plotted the IPR for various values of impurity density $\sigma$ and as a function of clustering probability $p$. We observe an increase in the IPR, indicating the degree of overall localization present in the states increases with an increase in clustering probability, $p$.

\section{Conclusion}
\label{sec:conclusions}
Motivated by the known positional correlations of impurities which arise in the configuration of impurities in materials, we adopted multiple approaches to investigate the electronic properties of a square lattice with nearest-neighbor hopping disorder.  In order to account for the clustering effect of impurities in a material, we used a semi-heuristic approach that introduces a clustering probability $p$ to determine the likelihood of having an impurity next to a given impurity.  Working with the impurity averaged T-matrix with this approach, we are able to obtain an important analytical result, Eq.\eqref{eq:analytic} that takes into account the impurity distributions with statistical information contained in Eq.\eqref{eq:Tcoeff}-Eq.\eqref{eq:Tcoeff_final}.

We found good agreement between the numerical and analytic approaches taken for the perturbative regime of hopping disorder. Particularly interesting effects are the suppression of quasiparticle lifetime, with single peaks originally at $\omega=0$ splitting into two sub-peaks with an increase in impurity clustering, as shown in Fig.~\ref{fig:tau}. 

Additionally, computing the inverse participation ratio demonstrates an increase in localization for more clustering of impurities, as seen in Fig.~\ref{fig:ipr}.  These combined numerical and analytical results which are expressed through the electron self-energy will appear in any quantity derived from the single-particle Green's function, including electrical and thermal transport, optical conductivity, and photoemission. (Note the transport and conductivity will depend on pairs of single-particle Green's functions where photoemission depends only on a single Green's function.)  Our work thus provides an important link between microscopic impurity distributions in a material and experimental observables \cite{PhysRevB.105.075425}.  This should help  elucidate the more complex effects of impurity clustering which are not possible to describe with simple isolated-impurity approximations.  

 \acknowledgements
 We gratefully acknowledge financial support from the
National Science Foundation through the Center for Dynamics and Control of Materials: an NSF MRSEC under
Cooperative Agreement No. DMR-1720595 and NSF DMR-2114825. M.V. gratefully acknowledges the support provided by the Deanship of Research Oversight and Coordination (DROC) at King Fahd University of Petroleum \& Minerals (KFUPM) for funding this work through exploratory research grant No. ER221002.

\bibliography{SENG}
\appendix
\section{Single-Impurity T-Matrix}
\label{sec:single}
For a single impurity at coordinate $\mathbf{R}$, the disorder component of the Hamiltonian can be written in the crystal momentum basis as,
\begin{equation}
\langle \mathbf{k} | \hat{H}_{imp} | \mathbf{k'} \rangle = -\frac{t'}{N} e^{i \mathbf{R}\cdot(\mathbf{k'} -\mathbf{k})}\left(\xi_\mathbf{k} + \xi_\mathbf{k'} \right).
\end{equation}

The n-fold scattering terms in the T-matrix series, when written as integrals over $n-1$ free crystal momenta, form a recurrence. Having evaluated the disorder average, each term becomes,
\begin{multline}
    T_n = \sigma (-t')^n \int \left(\prod_{i=1}^{n-1} \frac{d^2k_i}{(2\pi)^2} \right) ... \\ ...\left( (\xi_{\mathbf{k_0}}+\xi_\mathbf{k_1})\prod_{m=1}^{n-1} (\xi_\mathbf{k_m}+\xi_\mathbf{k_{m+1}})G^0_\mathbf{k_{m}} \right).
\end{multline}

We find all such integrals may be evaluated in terms of the integral of the undressed square lattice Green's function over the first Brillouin zone,
\begin{equation}
     g_{\square}(\omega) = \int_{-\pi}^{\pi}\int_{-\pi}^{\pi} \frac{d\mathbf{k}}{(2\pi)^2} \frac{1}{\omega +i\eta - \xi_\mathbf{k}} = \frac{2}{\pi \omega} \mathbf{K}(\frac{4t}{\omega}),
     \label{eq:sq}
\end{equation}
which is evaluated as a complete elliptic integral of the first kind, $\mathbf{K}$. The nth term in the T-matrix, diagonal in the crystal momentum, is proportional to $\alpha_n \xi^2_\mathbf{k} + (\beta_n + \gamma_n) \xi_\mathbf{k} +\delta_n $, where the coefficients satisfy the following system of recurrence equations,
\begin{subequations}
    \begin{align}
        \alpha_n = (\omega g_\square-1)\alpha_{n-1}+g_\square \gamma_{n-1}, \\
        \beta_n = (\omega g_\square-1)\beta_{n-1}+g_\square\delta_{n-1}, \\
        \gamma_n = (\omega g_\square-1)\gamma_{n-1} + \omega(\omega g_\square-1)\alpha_{n-1}, \\
        \delta_n = (\omega g_\square-1)\delta_{n-1}+ \omega(\omega g_\square-1)\beta_{n-1}.
    \end{align}
\end{subequations}

Using the easily computed values of these coefficients at $n=0,1$, exact expressions for the coefficients as a function of $n$ can be obtained with the help of Mathematica's RSolve. Finally summing these coefficients to $n=\infty$ yields the form presented in the manuscript.

\appendix
\end{document}